\RequirePackage{fix-cm}
\documentclass[smallextended]{svjour3}       
\smartqed  
\usepackage{graphicx}
\usepackage{nicefrac}
\usepackage{verbatim}
\usepackage{amsmath}
\usepackage{bm}
\usepackage{amssymb}
\usepackage{mathrsfs}

\newcommand{\eref}[1]{(\ref{#1})}
\newcommand{\eps }{\varepsilon }

\begin{document}

\title{Resonant scattering of light in a near-black-hole metric}

\author{Y. V. Stadnik\and
    G.~H.~Gossel \and
    V.~V.~Flambaum\and
    J.~C.~Berengut 
}

\institute{School of Physics, University of New South Wales, Sydney 2052, Australia  \\
\email{g.gossel@unsw.edu.au}
}

\date{Received: date / Accepted: date}

\maketitle

\begin{abstract}
We show that low-energy photon scattering from a body with radius $R$ slightly larger than its Schwarzschild radius $r_s$ resembles black-hole absorption.  This absorption occurs via capture to one of the many long-lived, densely packed resonances that populate the continuum. The lifetimes and density of these meta-stable states tend to infinity in the limit $r_s \to R$. We determine the energy averaged cross-section for particle capture into these resonances and show that it is equal to the absorption cross-section for a Schwarzschild black hole. Thus, a non-singular static metric may trap photons for arbitrarily long times, making it appear completely `black' before the actual formation of a black hole. 
\keywords{black hole \and resonant scattering \and photon scattering}
\PACS{04.62.+v, 04.70.Dy, 04.70.-s}
\end{abstract}
\section{Introduction}
In this work, we consider the scattering of photons by the gravitational field of a non-rotating, finite-sized body. Such finite-sized bodies have radius $R$ that slightly exceeds their Schwarzschild radius $r_s = 2GM/c^2$. We show that low-energy photon scattering from such objects resembles black hole absorption. The absorption arises due to the existence of a dense spectrum of narrow resonances (meta-stable states) whose lifetime and density tend to infinity in the limit $r_s \to R$.

Photons captured to such a resonant state are trapped on the interior of the body for a time $t \sim \hbar/\Gamma_n$, where $\Gamma_n$ is the width of a given resonance. For $r_s \to R$, both the energy spacing $D$ and width $\Gamma_n$ tend to zero (resonance lifetime $t \rightarrow \infty$), while their ratio remains finite. This allows us to define the total cross-section for particle capture into these long-lived states using the optical model~\cite{LLV3}, which is calculated by averaging over a small energy interval containing many resonances. At low energy the resonance capture cross-section is $\sigma_{a}=4\pi\eps^2 r_{s}^{4}/3$ where $\eps$ is the energy of the incident particle. The absence of a longitudinal mode for photons means there is no state with total angular momentum equal to zero, hence the above cross-section tending to zero for zero energy.

This cross-section exactly matches the absorption cross-section for a Schwarz\-schild black hole calculated by assuming complete absorption at the event horizon (see Refs.~\cite{Matzner,Starob,Unruh,Sanchez,Das,Crispino,Decanini}) derived previously for massless spin-1 particles~\cite{Fabbri,Kanti}. Note, however, that our calculation does not impose any special conditions at the boundary. Thus by considering the purely elastic cross-section of low-energy incident photons, we find that the absorption properties of a body with $r_s \rightarrow R$ resemble those of a black hole. 

The gravitational field of these near-black-hole objects is described using a suitable metric to model the interior, which is then joined to the standard Schwarzschild exterior metric at the boundary of the body. The above result is shown to be valid for any interior metric satisfying the following conditions: continuity with the Schwarzschild exterior at $r=R$, a potential that deepens and develops a singularity in the black hole limit (allowing the particle to be treated semi-classically) and that this potential can be modeled as a harmonic oscillator in some finite region around the origin. We expect these conditions to be met by a large class of metrics; in this paper we present two metrics for which these conditions hold. 
 
Additionally, the results presented include the spin-0 (scalar) case considered previously (for $j=0$) \cite{Flambaum2012}, but given here for arbitrary angular momentum $j$. In this case taking $r_s \to R$ once again yields the black hole absorption cross-section given in \cite{Unruh}.

For a more detailed discussion of previous calculations involving scattering of various particles in the black hole spacetime we turn the readers attention to the introductions in \cite{Flambaum2012,Crispino} and references therein. 
As in previous work \cite{Flambaum2012} we perform both analytical and numerical calculations, with good agreement between the two.
\section{Wave equations}
The Klein-Gordon equation for a massless spin-0 particle on a curved manifold (with $\hbar =c=1$) reads
\begin{equation}
\label{KG_massless}
\partial_\mu (\sqrt{-g} g^{\mu\nu}\partial_\nu \Psi) = 0 .
\end{equation}
For massless spin-1 particles (photons) Maxwell's equations on a curved manifold free of charges read
\begin{equation}
\label{Maxwell_eqns}
\partial_{\beta}(\sqrt{-g}F^{\alpha \beta}) = 0,
\end{equation}
where $F^{\alpha \beta}=\partial^\alpha A^\beta-\partial^\beta A^\alpha$ and $A^\mu$ is the contravariant electromagnetic 4-potential. One may write a general static, spherically symmetric metric in the form
\begin{equation}
\label{General_metric}
ds^{2}= - e^{\nu \left( r \right)}dt^{2} + e^{\lambda \left( r \right)}dr^{2} + r^{2}d\Omega^{2} .
\end{equation}
By applying the separation of variables 
\begin{gather}
\Psi = e^{-i\eps t}\phi(r)\, Y_{jm}(\theta ,\varphi ),\nonumber \\
A_\varphi = e^{-i \eps t} r \phi(r) \sin\theta\, \frac{d P_j \left(\cos\theta\right)}{d\theta} \nonumber
\end{gather}
to (\ref{KG_massless}) and (\ref{Maxwell_eqns}) respectively in the metric (\ref{General_metric}), one arrives at the radial differential equations for spin-0 and spin-1 respectively. In the latter we have implemented a gauge such that $A_\varphi$ is the only non-zero component \cite{Wheeler}.

Defining $s$ as the particle spin, either 0 or 1, the radial equations for both spins can be combined to yield 
\begin{equation}
\label{eq:GenEqn}
\frac{d^{2}\phi}{dr^{2}} + \left[\left(\frac{\nu ' - \lambda '}{2}\right) + \frac{2}{r}\right]\frac{d\phi}{dr} + \left[e^{\lambda - \nu}\varepsilon ^{2} + s^2 \left(\frac{\nu ' - \lambda '}{2r}\right) - \frac{j(j+1)e^{\lambda}}{r^{2}}\right]\phi (r) = 0 ,
\end{equation}
where $j$ the total angular momentum such that $j-s \ge 0$.
\section{Interior Solution}
\label{sec:gen_int_soln}
Equation~\eref{eq:GenEqn} can be transformed into a Schr\"odinger-like equation by making the substitution $\phi(r) = \chi(r)/r$ and then mapping the radial coordinate to the Regge-Wheeler ``tortoise''
coordinate $r^*$ defined by $dr^{*}=e^{(\lambda - \nu)/2}dr$. This gives us
\begin{equation}
\frac{d^2\chi(r^*)}{d(r^*)^2} + \left[\varepsilon^{2} +\frac{(s^2-1)}{2r}\frac{d}{dr}e^{\nu(r)-\lambda(r)}- \frac{j(j+1)e^{\nu (r)}}{r^{2}} \right]\chi(r^{*}) = 0 .
\label{eq:TortEqn}
\end{equation}
In the subsections below we construct a solution to the above equation by dividing the interior into two regions. This solution is valid under certain constraints placed on the metric coefficients $e^\nu$ and $e^\lambda$. The specific metrics considered in later sections are entirely consistent with the restrictions imposed. It is worth noting that in the s-wave scalar case $s = j = 0$ there is only one interior region akin to the one defined in Sec.~\eref{sec:R2I}, albeit with different constraints. As such, this case is not treated here and we instead refer the reader to our previous treatment of this case in \cite{Flambaum2012}. For the purposes of the discussion below we note that for the interior metric of near-black-hole object, $e^{\nu \left( r \right)} \to 0$ for $0\leq r\leq R$, as time slows down in the limit $r_s \to R$.
\subsection{Interior Region I}
\label{sec:R1I}
In the vicinity of the origin spacetime, and indeed the potential, must be locally flat. This necessitates that both $e^\nu(r)$ and $e^\lambda(r)$ be approximately constant, and furthermore $e^\lambda(r) \approx 1$, in the area around the origin. Under these conditions we may ignore the second term in square brackets in Eqn.~\eref{eq:TortEqn} (moreover, it is zero for spin $s=1$), and we may approximate the tortoise coordinate by $r^{*}_{0} \approx e^{-\nu_0/2}r$, where $\nu_0$ is a constant. 

For the metrics we consider, the second bracketed term in Eqn.~\eref{eq:TortEqn} is always smaller than either the centrifugal term or $\eps^2$ throughout Region I. Therefore in Region I we re-write Eqn.~\eref{eq:TortEqn} as
\begin{equation}
\label{eq:TortEqnNew}
\chi_{\mathrm{I}}''(r^{*}) + \left[\varepsilon^{2} - \frac{j(j+1)}{(r^{*})^{2}} \right]\chi_{\mathrm{I}}(r^{*}) = 0,
\end{equation}
the regular solution of which is the Bessel function
\begin{equation}
\label{eq:BesselSoln}
\chi_{\mathrm{I}}(r^*) = A_\mathrm{I} \sqrt{\eps r^*} J_{j+1/2} (\eps r^*).
\end{equation} 
\subsection{Interior Region II}
\label{sec:R2I}
This region is defined as the area of the interior where the energy term $\eps^2$ dominates over other terms in square brackets in Eqn.~\eref{eq:TortEqn}. Note that these terms are suppressed by a factor of $e^{\nu}$, which tends to zero in the black hole limit. Therefore the wave equation is simply
\begin{equation}
\chi_{\mathrm{II}}''(r^{*}) +\eps^2 \chi_{\mathrm{II}}(r^{*})=0,
\end{equation}
which has the solution
\begin{gather}
\label{eq:FreeSol}
\chi_{\mathrm{II}}(r^{*})  =A_{\mathrm{II}} \sin\left(\eps r^*+\phi \right), \\ \notag{}
\implies \chi_{\mathrm{II}}(r) =A_{\mathrm{II}} \sin\left(\eps \int_{0}^{r} e^{[\lambda(r')-\nu(r')]/2}dr' + \phi \right),
\end{gather}
where $\phi$ is a phase to be determined by matching to the solution in Region I in an appropriate overlap region. For the metrics we consider, Region II is valid up to a point near the boundary beyond which $\eps^2$ is no longer the dominant term. However, for the metrics we consider it may be shown that in the black hole limit the size of this region decreases as a function of $R-r_s$ such that it does not appreciably alter the phase of the wavefunction from Region II to the boundary $r=R$.
\subsection{Matching in the overlap region}
 It can be shown that for both the Florides and Soffel metrics there exists an overlap region where the combined conditions defining regions I and II are satisfied. This overlap region exists for $r$ such that:
\begin{align}
&r^* \approx e^{-\nu_0/2} r \notag{}, \\
&\frac{(s^2-1)}{2r}\frac{d}{dr}e^{\nu(r)-\lambda(r)} \ll \eps^2, \notag{} \\
&\frac{j(j+1)e^{\nu (r)}}{r^{2}}  \ll \eps^2.
\label{eq:Condition1}
\end{align}
Lastly, the condition
\begin{equation}
\label{eq:Condition2}
\eps r^*= \eps r e^{-\nu_0/2} \gg 1,
\end{equation}
allows us to take the asymptotic form of the solution in Region I given by \eref{eq:BesselSoln}, $\sqrt{\eps r^*} J_{j+1/2} (\eps r^*) \to \sin(\eps r^*-j\pi/2)$. Matching to this solution in the overlap region defined above gives the final Region II solution as 
\begin{equation}
\label{eq:IntSolnFinal}
\chi_\mathrm{II}(r) =A_\mathrm{I} \sin\left[ \Phi(r) -j\pi/2 \right],
\end{equation}
where
\begin{equation}
\label{eq:PhaseDef}
\Phi(r) =\eps \int_{0}^{r} e^{[\lambda(r')-\nu(r')]/2}dr'.
\end{equation}
Furthermore, at the boundary we have $\Phi'(R) = \left.\eps e^{(\lambda-\nu)/2}\right|_R =\eps R/(R-r_s)$. The latter is computed by imposing continuity of the metric at the boundary (where all but the energy term in Eqns.~\eref{eq:GenEqn} and \eref{eq:TortEqn} are suppressed). Defining 
\begin{equation}
\label{eq:LambdaDef}
\Lambda(r_s) = \int_{0}^{R}e^{(\lambda-\nu)/2}dr , 
\end{equation}
the logarithmic derivative of the interior wavefunction at $r=R$ can be written as
\begin{equation}
\label{eq:IntLD}
\left. \frac{\phi'(r)}{\phi(r)}\right|_{R} = \frac{\eps R}{R-r_s}\cot\left[\eps \Lambda(r_s) -j \pi/2 \right]-\frac{1}{R}.
\end{equation}
In the black-hole limit $e^\nu \to 0$, therefore from Eqn.~\eref{eq:LambdaDef} we see that as $r_s\to R$, $\Lambda(r_s)$ tends to infinity. This is because $\Lambda(r_s)$ is related (but not equal) to the total phase accumulated by the particle on the interior. This phase is large due to the wave function oscillating many times on the interior as the particle moves rapidly in the strong field. However this integral also gives the classical time that a massless particle ($ds^2=0$) spends on the interior, which goes to infinity in the black hole limit.
\section{Exterior Solution}
The metric on the exterior ($r\geq R$) of a static massive body is given by the standard Schwarzschild metric, which yields the following radial differential equation
\begin{equation}
\label{eq:ExtEqn}
\frac{d^{2}\phi}{dr^{2}} + \left(\frac{1}{r-r_{s}} + \frac{1}{r}\right)\frac{d\phi}{dr} + \left[\frac{\varepsilon^{2}r^{2}}{\left(r-r_{s}\right)^{2}} + \frac{s^2r_{s}}{r^{2}\left(r-r_{s}\right)} - \frac{j(j+1)}{r(r-r_{s})} \right]\phi(r) = 0 .
\end{equation}
\subsection{Exterior Region I}
For $\eps \ll \sqrt{r-r_s}$ (later verified for resonance energies $\eps_n$), Eqn.~\eref{eq:ExtEqn} becomes
\begin{equation}
\label{Radial_DE_ExtII}
\frac{d^{2}\phi}{dr^{2}} + \left(\frac{1}{r-r_{s}} + \frac{1}{r}\right)\frac{d\phi}{dr} + \left[\frac{s^2r_{s}}{r^{2}\left(r-r_{s}\right)} - \frac{j(j+1)}{r(r-r_{s})} \right]\phi(r) = 0 , 
\end{equation}
which has the exact solution
\begin{align}
\label{E2soln}
&\phi_{\textrm{I}}(r) = \alpha_{1}\left(\frac{r}{r_{s}}\right)^{s} P_{j-s}^{(2s,0)} \left(1-\frac{2r}{r_{s}}\right)  \\
&+ \beta_{1} \left(\frac{r_{s}}{r}\right)^{j+1} \phantom{e}_{2}F_{1}\left(j-s+1,j+s+1,2j+2;\frac{r_{s}}{r}\right) ,\notag{}
\end{align}
where $P_{n}^{(a,b)}(x)$ represent the Jacobi polynomials and $_{2}F_{1}(a,b,c;z)$ is the Gaussian hypergeometric function. 
This solution is valid from the boundary $r=R$ (provided $\eps \ll \sqrt{R-r_s}$) and while $\eps^2 r^2 \ll j (j+1)$.  
\subsection{Exterior Region II}
Taking $r\gg r_s$ in Eqn. \eref{eq:ExtEqn} we make the substitutions $\phi(r)=\chi(r)/r$ and $r = \rho / \varepsilon$. This yields
\begin{equation}
\label{Coulomb_eqn_std}
\frac{d^{2}\chi(\rho)}{d\rho^{2}} + \left[1 + \frac{2\varepsilon r_{s}}{\rho} - \frac{j(j+1)}{\rho^{2}} \right]\chi(\rho) = 0 .
\end{equation}
The corresponding solution in terms of the regular and irregular Coulomb wave functions~\cite{AbrmSteg} reads
\begin{equation}
\label{E3soln}
\phi_{\textrm{II}}(r) = \frac{\alpha_{2}F_{j}(\varepsilon r_{s},\varepsilon r) + \beta_{2}G_{j}(\varepsilon r_{s},\varepsilon r)}{r} .
\end{equation}
There exists an overlap between the two regions described above when $r \gg r_s$  but $\eps^2 r^2 \ll j(j+1)$ (which automatically satisfies $\eps \ll \sqrt{r-r_s}$). In this overlap region we can use the asymptotic form ($\eps r \ll 1$) of the Coulomb wavefunctions in~\eref{E3soln}. Matching with \eref{E2soln} we arrive at the following relationship between the coefficients of the solutions in the two exterior regions:
\begin{gather}
\label{AIII/BIII}
\frac{\alpha_{2}}{\beta_{2}} =
  \frac{\alpha_1}{\beta_1} \frac{(2j)! (-1)^{j-s}}{(j-s)!(j+s)! C^2 (2j+1) (\varepsilon r_s)^{2j+1}} ,\\
C=\frac{2^j e^{\varepsilon r_s \pi /2}|\mathrm{\Gamma}\left(j+1-i \varepsilon r_s \right)|}{(2j+1)!}. \nonumber
\end{gather}
\section{S-Matrix}
The solution to Eqn.~\eref{eq:ExtEqn} at large distances can also be written in terms of outgoing and incoming waves as 
\begin{equation}
\label{Coulomb_scatter}
\phi_{\mathrm{II}} (r)= \frac{A_j e^{i z} +B_j e^{-iz}}{r} ,
\end{equation}
where $z = \varepsilon r+ \varepsilon r_s \ln(2\varepsilon r) + \delta_{j}^{C}  - j\pi/2$ and $\delta_{j}^{C}~=~ \arg\left[\mathrm{\Gamma}\left(j + 1 + i\varepsilon r_{s}\right)\right]$ is the Coulomb phase shift. This allows us to write the scattering matrix as~\cite{LLV3}
\begin{equation}
S_j = (-1)^{j+1} \frac{A_j}{B_j}e^{2i\delta_{j}^{C}} .
\end{equation}
Imposing the condition $\varepsilon r \gg 1$ in Eqn.~\eref{E3soln}, we have the following asymptotic form of the wavefunction in region~II
\begin{equation}
\label{E3soln_as_r->inf}
\phi_{\textrm{II}}(r) = \frac{\alpha_2 \sin(z) + \beta_2 \cos(z)}{r} ,
\end{equation}
which, upon matching to (\ref{Coulomb_scatter}), gives the scattering matrix as
\begin{equation}
\label{S-matrix}
S_j = - \frac{1-\frac{i\alpha_2}{\beta_2}}{1+\frac{i\alpha_2}{\beta_2}}   e^{2i\delta_{j}^{C}} .
\end{equation}
Note that $\delta_{j}^{C}$ is small compared to the total phase accumulated on the interior, given by $\Lambda(r_s)$, and slowly varying for $\eps \ll 1$.
\section{Matching of wavefunctions at boundary and resonance energies}
Matching the logarithmic derivatives of the exterior~\eref{E2soln} and interior~\eref{eq:IntLD} wavefunctions at $r=R$ and taking the black hole limit $r_s \to R$ gives
\begin{equation}
\label{eq:coeff-A2/B2_simplified}
\frac{\alpha_1}{\beta_1}
 = -\frac{(-1)^{s-j}\mathrm{\Gamma}(2j+2)}{\mathrm{\Gamma}(j-s+1)\mathrm{\Gamma}(j+s+1)}\,
   \frac{\tan\left(\eps \Lambda(r_s)-j\pi/2 \right)}{\eps R}.
\end{equation}
Resonances occur at energies where the absorption cross-section is maximized, i.e.~$S=-1$. This is achieved in Eqn.~\eref{S-matrix} when $\alpha_2/\beta_2 = 0$, which by Eqn. (\ref{AIII/BIII}) is equivalent to ${\alpha_1}/{\beta_1}=0$. Setting ${\alpha_1}/{\beta_1}$ to zero in \eref{eq:coeff-A2/B2_simplified} results in the resonance condition for the energy
\begin{equation}
\label{eq:ResEn}
\varepsilon_{n} = \frac{n\pi + j\frac{\pi}{2}}{ \Lambda(r_s) } ,
\end{equation}
where $n = 1, 2, \dots$. Note that these resonance energies strongly depend on the value of $j$, but in this approximation do not depend on $s$.
\section{Resonance Widths}
The full resonance is obtained by extending $\eps$ into the complex plane. The scattering matrix has a pole at complex energy $\eps = \eps_n -i\Gamma_n/2$ which corresponds to the resonance condition
\begin{equation}
\label{complex_poles}
1+\frac{i \alpha_2}{\beta_2} = 0.
\end{equation}
We may also express $\alpha_1/\beta_1$ in the vicinity of a resonance as
\begin{align}
\label{eq:Taylor_exp_energy}
\frac{\alpha_1}{\beta_1} (\varepsilon) 
 &= \frac{\alpha_1}{\beta_1} (\eps_n) +\left[ \frac{\partial}{\partial \varepsilon}\left.\left(\frac{\alpha_1}{\beta_1}\right)\right]\right|_{\varepsilon = \varepsilon_{n}}(\varepsilon - \varepsilon_{n}).
\end{align}
As detailed previously, on resonance $\alpha_1/\beta_1=0$. At the complex pole of the scattering matrix, \mbox{$\eps =\eps_n -i\Gamma_n/2$}, this gives
\begin{equation}
\label{eq:Taylor_exp_energyFinal}
\frac{\alpha_1}{\beta_1} = \left[\frac{\partial}{\partial \varepsilon}\left.\left(\frac{\alpha_1}{\beta_1}\right)\right]\right|_{\varepsilon = \varepsilon_{n}} \left(\frac{-i \Gamma_{n}}{2}\right).
\end{equation}
Let $\alpha_2/\beta_2 = f(\eps) \alpha_1/\beta_1$ [Eqn.~\eref{AIII/BIII}], then using the above expression we may write Eqn.~\eref{complex_poles} as
\begin{equation}
\label{complex_poles2}
1+f(\eps_n) \frac{\Gamma_n}{2}\left.\frac{\partial}{\partial \eps}\left(\frac{\alpha_1}{\beta_1}\right)\right|_{\eps=\eps_n} =0 ,
\end{equation}
Taking the derivative of $\alpha_1/\beta_1$ given by Eqn.~\eref{eq:coeff-A2/B2_simplified} and solving for $\Gamma_n$ in \eref{complex_poles2} gives the resonance widths
\begin{equation}
\label{eq:WidthEqn}
\Gamma_{n} = \frac{2C^2 (\varepsilon R)^{2j+2}((j-s)!)^2 ((j+s)!)^2}{((2j)!)^2 \Lambda(r_s) },
\end{equation}
where we have assumed $\eps_n \Lambda(r_s)\gg j \pi/2$, i.e. large $n$ (see \eref{eq:ResEn}), which is already assumed when deriving the interior solution.
\section{Absorption cross-section}
As discussed in Section~\ref{sec:gen_int_soln}, in the black hole limit we find that $\Lambda(r_s)$ tends to infinity. Thus by Eqns.~\eref{eq:ResEn} and \eref{eq:WidthEqn} both $\varepsilon_{n}$ and $\Gamma_{n}$ tend to zero in the limit $r_{s} \rightarrow R$ for any fixed, finite values of $n$ and $j$. However, the ratio $\Gamma_n/D$ remains constant, where $D=\eps_{n+1}-\eps_n\simeq \pi/\Lambda(r_s)$ is the spacing  between adjacent levels. This allows us to use the optical-model (energy-averaged absorption cross-section) \cite{LLV3}. This is obtained by averaging over a small energy interval containing many resonances and reads
\begin{equation}
\label{opt_model_x-sect}
\bar{\sigma}_{a}^{\mathrm{opt}} = \sum\limits_{j-s=0}^{\infty} \frac{2\pi^{2}}{\varepsilon^{2}} \frac{\Gamma_{n}}{D} (2j+1).
\end{equation}
Substituting \eref{eq:ResEn} and \eref{eq:WidthEqn} into (\ref{opt_model_x-sect}), gives
\begin{equation}
\label{cross-section}
\bar{\sigma}_{a}^{\mathrm{opt}} = \sum\limits_{j-s=0}^{\infty} \frac{4 \pi (2j+1)C^2 \varepsilon^{2j} R ^{2j+2}((j-s)!)^2 ((j+s)!)^2}{((2j)!)^2 } ,
\end{equation}
which is independent of $\Lambda(r_s)$ and thus of the interior metric.

Therefore, in the low energy limit the cross-sections for massless scalar particles and photons are
\begin{center}
     \begin{tabular}{l | c  c  }
   &  Spin 0 &   Spin 1\\ [.5ex] \hline \\ [-2ex]
\( \displaystyle\bar{\sigma}_{a}^{\mathrm{opt}}\) 
&   \( \displaystyle4\pi r_{s}^{2}\)
& \( \displaystyle\frac{4\pi r_{s}^{4}\epsilon^2}{3}\) 
     \end{tabular}
\end{center}
The above expressions exactly match the cross sections for these particles incident on a Schwarzschild black hole \cite{Unruh,Fabbri,Kanti}.
\section{Specific interior metrics}
\label{sec:Specifics}
In this section we present calculations involving two specific interior metrics that allow the $r_s \rightarrow R$ limit to be taken: the Florides \cite{Florides74} and Soffel \cite{Soffel77} metrics. Specifically, we verify our analytic solutions with numerically calculated resonance widths and energies via the short range phase shift  $\delta(\eps)$. To calculate $\delta(\eps)$ we solve the second-order differential equation (\ref{eq:GenEqn}) numerically, for given $e^\nu$ and $e^\lambda$, with the boundary condition $\phi(r\to0) \sim r^j$ using {\em Mathematica} \cite{math}. This solution provides a real boundary condition for the {\em exterior} wave function at $r=R$. (We set $R=1$ in the numerical calculations). Equation \eref{eq:ExtEqn} is then integrated outwards to large distances $r\gg r_s$. In this region Eq.~(\ref{eq:ExtEqn}) takes the form of a non-relativistic Schr\"odinger equation for a particle with momentum $\eps $ and unit mass in the Coulomb potential with charge $Z=-r_s\eps ^2$. Hence, we match the solution with the asymptotic form \cite{LLV3}
\begin{equation}\label{eq:CoulombMatch}
\chi(r)\propto \sin [\eps r - (Z/\eps ) \ln 2 \eps r +\delta_C+ \delta-j \pi/2 ]
\end{equation}
and determine the short-range (numeric) phase shift $\delta $. It is found that this phase possesses steps of height $\pi$ at the resonance positions $\eps_n$. We fit the step profile of an individual resonance to the  Breit-Wigner function
\begin{equation}
\delta(\eps\simeq\eps_n)= \delta_n + \arctan\left[\frac{\eps-\eps_n}{\Gamma_n/2}\right]
\end{equation}
where $\delta_n$ is a constant, from which
we extract the numeric resonance widths and positions $\Gamma_n$ and $\eps_n$.
\subsection{Florides Interior}
\begin{figure}[tb]
\begin{center} 
\includegraphics[width=1\textwidth]{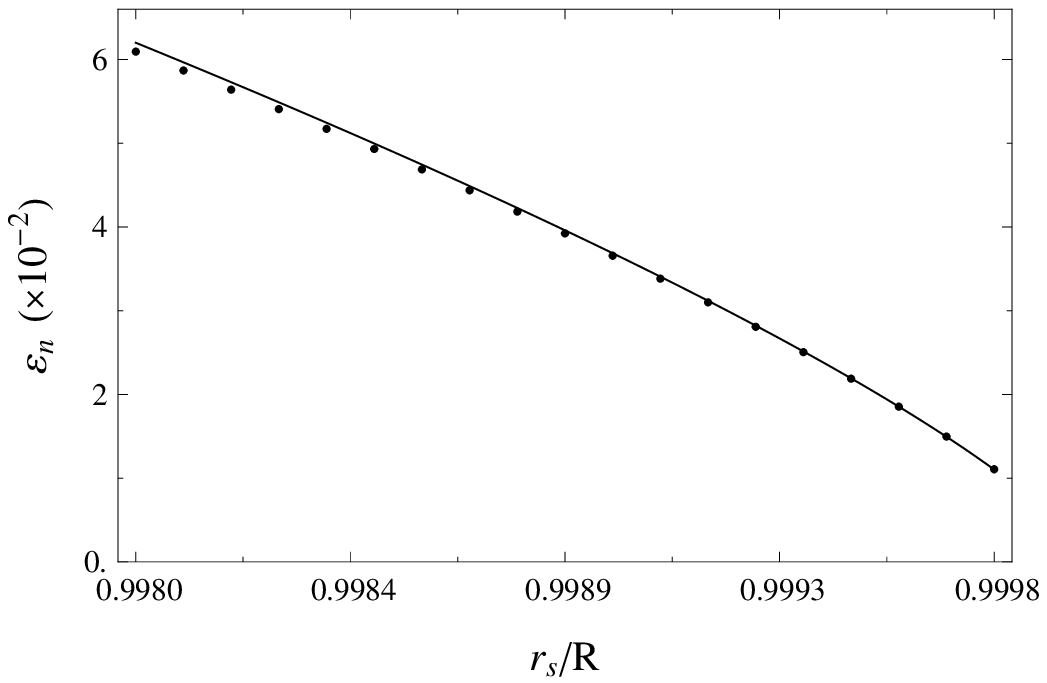}
\caption{Energies of the $n=2$ resonance in the Florides metric with $s=j=1$. Closed circles  indicate numeric data, the solid line indicates analytic $\eps_n$ given by Eqn.~\eref{eq:ResEn} with $\Lambda(r_s)$ given by Eqn.~\eref{eq:FloridesL}.
\label{fig:FloridesEn}}
\end{center}
\end{figure}

\begin{figure}[tb]
\begin{center} 
\includegraphics[width=1\textwidth]{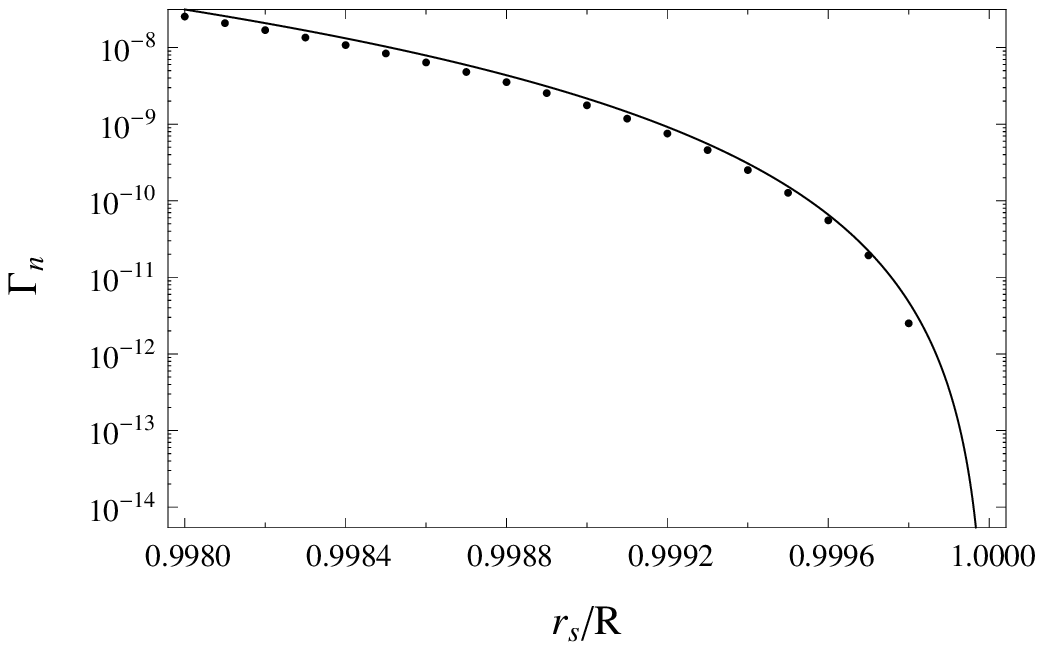}
\caption{Widths of the $n=2$ resonance in the Florides metric with $s=j=1$. Closed circles indicate numeric data, the solid line indicates analytic $\Gamma_n$ given by Eqn.~\eref{eq:WidthEqn} with $\Lambda(r_s)$ given by Eqn.~\eref{eq:FloridesL}.
\label{fig:FloridesW}}
\end{center}
\end{figure}

The Florides metric is characterized by
\begin{equation}
\label{eq:FInt}
e^{\nu(r)}_{\mathrm{F}}=\frac{(1-r_s/R)^{3/2}}{\sqrt{1-r_sr^2/R^3}},\quad
e^{\lambda(r)}=\left(1-\frac{r_s r^2 }{R^3}\right)^{-1}.
\end{equation}
The leading term of $\Lambda(r_s)$ is
\begin{align}
\label{eq:FloridesL}
\Lambda_\textrm{F}(r_s) &\overset{r_s \rightarrow R}{=} \frac{\pi^{3/2} R}{\sqrt{2} \Gamma(\nicefrac{1}{4}) \Gamma(\nicefrac{5}{4}) (1-r_s /R)^{3/4}},\notag{} \\
 &\ \ \approx \ 1.198\, (1-r_s/R)^{-3/4}.
\end{align}
The resulting resonance energies and widths are compared with their numeric counterparts in Figures \eref{fig:FloridesEn} and \eref{fig:FloridesW} respectively.

\subsection{Soffel Interior}
\begin{figure}[!tb]
\begin{center} 
\includegraphics[width=1\textwidth]{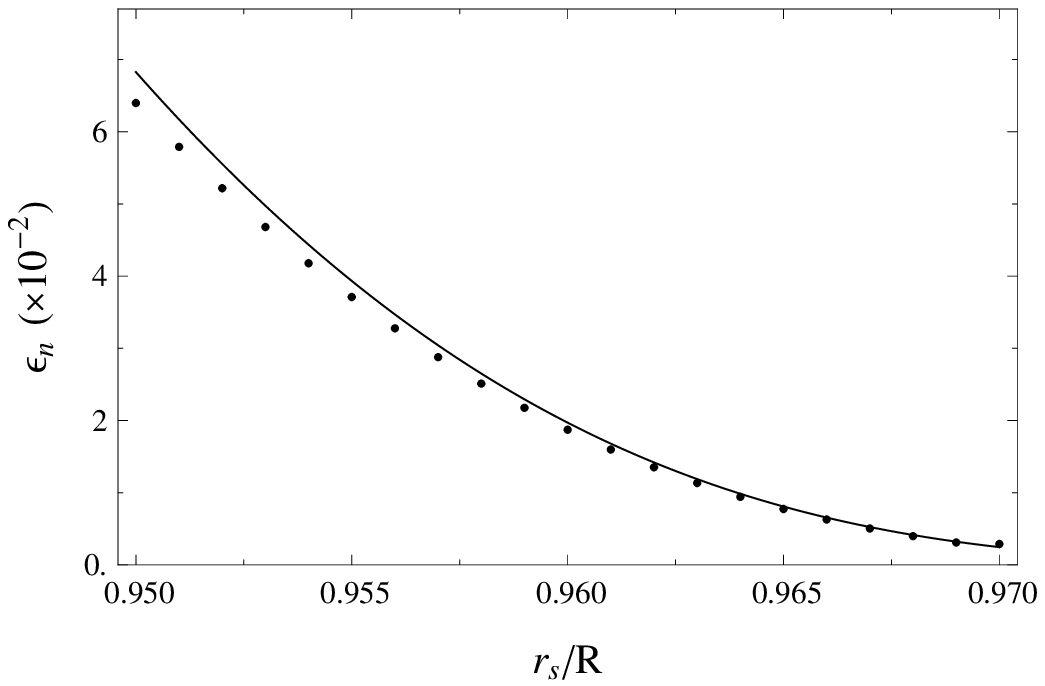}
\caption{Energies of the $n=4$ resonance in the Soffel metric with $s=j=1$. Closed circles  indicate numeric data, the solid line indicates analytic $\eps_n$ given by Eqn.~\eref{eq:ResEn} with $\Lambda(r_s)$ given by Eqn.~\eref{eq:SoffelL}.
\label{fig:SoffelEn}}
\end{center}
\end{figure}

\begin{figure}[tb]
\begin{center} 
\includegraphics[width=1\textwidth]{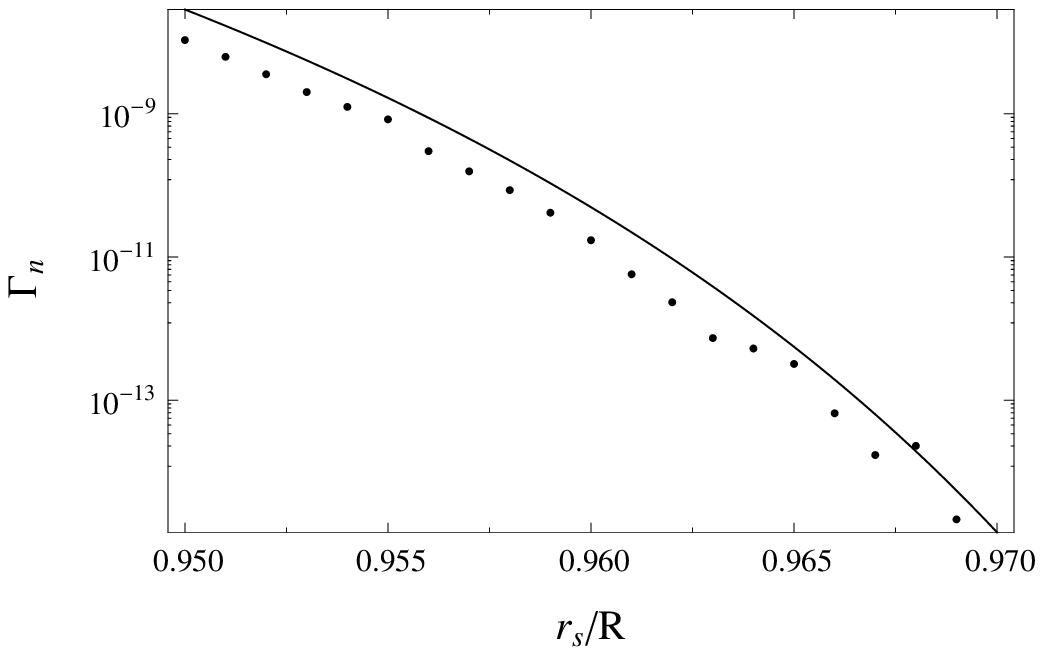}
\caption{Widths of the $n=4$ resonance in the Soffel metric with $s=j=1$. Closed circles  indicate numeric data, the solid line indicates analytic $\Gamma_n$ given by Eqn.~\eref{eq:WidthEqn} with $\Lambda(r_s)$ given by Eqn.~\eref{eq:SoffelL}.
\label{fig:SoffelWidths}}
\end{center}
\end{figure}

The Soffel metric is characterized by \cite{Soffel77}
\begin{equation}
\label{eq:SInt}
e_\textrm{So}^{\nu(r)}= \left(1-\frac{r_s}{R}\right) 
\exp \left[-\frac{r_s (1-r^2/R^2)}{2R (1-r_s/R)}\right],
\end{equation}
with $e^{\lambda(r)}$ equal to that of the Florides case. The leading term of $\Lambda(r_s)$
\begin{equation}
\label{eq:SoffelL}
\Lambda_\textrm{So}(r_s) \overset{r_s \rightarrow R}{=}R\sqrt{\pi}\exp\left[\frac{r_s/R}{4(1-r_s/R)}\right].
\end{equation}
Analytic and numeric $\varepsilon_n$ and $\Gamma_n$ for the Soffel metric are compared in Figures \eref{fig:SoffelEn} and \eref{fig:SoffelWidths} respectively. The widths in Fig.~\eref{fig:SoffelWidths} illustrate that the exponential suppression gives rise to numerical instabilities as $r_s \to R$ (and thus why the $r_s/R$ values used in the Soffel case are much lower than those in the Florides case).

\section{Conclusions}
The problems of the scattering of low-energy, massless spin-0 and spin-1 particles from a massive, static, spherical body have been considered. We have shown that such scattering is characterized by a dense set of long lived resonances. Capture to these long-lived states gives rise to effective absorption in a purely potential scattering setting. In the black hole limit the cross-section for absorption exactly equals the cross-section in the pure black hole case (for low energy). Thus scattering of photons (and massless scalars) by a near-black-hole object resembles black hole absorption.

We thank G. F. Gribakin for useful discussions. This work is supported by the Australian Research Council.


\begin{thebibliography}{99} 
\bibitem{LLV3} L.~D.~Landau and E.~M.~Lifshitz, \textit{Quantum Mechanics},
3rd Ed. (Butterworth-Heinemann, Oxford, 1977).

\bibitem{Matzner}R. A. Matzner, \textit{ 		
 Scattering of Massless Scalar Waves by a Schwarzschild ``Singularity''}, J. Math. Phys. {\bf 9}, 163 (1968).

\bibitem{Starob}A. A. Starobinskii, \textit{Amplification of waves during reflection from a rotating black hole}, Zh. Eksp. Teor. Fiz. {\bf 64}, 48 (1973)
[Sov. Phys. JETP {\bf 37}, 28 (1973)].

\bibitem{Unruh} W.~G.~Unruh, \textit{Absorption cross section of small black holes}, Phys. Rev. D {\bf 14}, 3251 (1976); Thesis, Princeton Univ., 1971 (unpublished).

\bibitem{Sanchez}N. Sanchez, \textit{Absorption and emission spectra of a Schwarzschild black hole}, Phys. Rev. D {\bf 18}, 1030 (1978).

\bibitem{Das}S. R. Das, G. Gibbons, and S. D. Mathur, \textit{Universality of Low Energy Absorption Cross Sections for Black Holes},  Phys. Rev. Lett.
{\bf 78}, 417 (1997).

\bibitem{Crispino}L. C. B. Crispino, S. R. Dolan, and E. S. Oliveira, \textit{Electromagnetic wave scattering by Schwarzschild black holes}, 
Phys. Rev. Lett. {\bf 102}, 231103 (2009).

\bibitem{Decanini}Y. D\'ecanini, G. Esposito-Far\`ese, A. Folacci, \textit{Universality of high-energy absorption cross sections for black holes}, 
Phys. Rev. D {\bf 83}, 044032 (2011).

\bibitem{Fabbri} R.~Fabbri, \textit{Scattering and absorption of electromagnetic waves by a Schwarzschild black hole},  Phys. Rev. D \textbf{12}, 933, (1975).

\bibitem{Kanti} P.~Kanti and J.~March-Russell, \textit{Calculable corrections to brane black hole decay. II. Greybody factors for spin 1/2 and 1}, Phys. Rev. D \textbf{67}, 104019, (2003).

\bibitem{Flambaum2012}
V.~V.~Flambaum, G.~H.~Gossel and G.~F.~Gribakin, \textit{Dense spectrum of resonances and particle capture in a near-black-hole metric},  Phys. Rev. D {\bf 85}, 084027
(2012).

\bibitem{Wheeler}
J.~A.~Wheeler. \textit{Geons}, Phys. Rev., {\bf97}, 511, (1955).

\bibitem{AbrmSteg}
M.~Abramowitz and I.~A.~Stegun, \textit{Handbook of Mathematical Functions with Formulas, Graphs, and Mathematical Tables}, 10th Ed. (National Bureau of Standards, 1972).

\bibitem{Florides74}  P.~S.~Florides, \textit{A new interior Schwarzschild solution}, Proc. R. Soc. Lond. A {\bf 337}, 529 (1974).

\bibitem{Soffel77}M.~Soffel, B.~M\"uller, and W.~Greiner, \textit{Particles in a stationary spherically symmetric gravitational field}, 
J. Phys. A {\bf 10}, 551 (1977).

\bibitem{math}{\em Mathematica, Version 7.0}
(Wolfram Research, Inc., Champaign, IL, 2008).

\end{thebibliography}
\end{document}